\documentclass[preprint,aps,12pt,preprintnumbers,eqsecnum,nofootinbib]{revtex4}
\usepackage{graphicx}
\usepackage{subfigure}
\usepackage{epstopdf} 
\usepackage{braket}
\usepackage{amssymb,amsmath,amsfonts,comment,latexsym,graphicx,epstopdf,bbold,float}
\usepackage{color}
\usepackage{fancyvrb}
\usepackage{chngcntr}
\usepackage{etoolbox}
\usepackage{bbm,epsfig}
\usepackage{array}
\usepackage{hepunits}

\newcommand{\be}{\begin{equation}}
\newcommand{\ee}{\end{equation}}
\newcommand{\ba}{\begin{eqnarray}}
\newcommand{\ea}{\end{eqnarray}}

\newcommand{\grts}{\raise.3ex\hbox{$>$\kern-.75em\lower1ex\hbox{$\sim$}}}
\newcommand{\lets}{\raise.3ex\hbox{$<$\kern-.75em\lower1ex\hbox{$\sim$}}}

\def\braket#1{\mathinner{\langle{#1}\rangle}}

{\catcode`\|=\active\gdef\Braket#1{\left<\mathcode`\|"8000\let|\bravert {#1}\right>}}

\def\bravert{\egroup\,\vrule\,\bgroup}

\usepackage{subfigure}

\usepackage{color}
\usepackage{amssymb,amsmath,amsfonts}
\usepackage{epstopdf} 
\usepackage{braket}

\begin{document}
%
%
\title{\vspace*{0.5in} 
Flavor from the double tetrahedral group without supersymmetry: flavorful axions and neutrinos
\vskip 0.1in}
\author{Christopher D. Carone}\email[]{cdcaro@wm.edu}
\author{Marco Merchand}\email[]{mamerchandmedi@email.wm.edu}

\affiliation{High Energy Theory Group, Department of Physics,
College of William and Mary, Williamsburg, VA 23187-8795}

\date{\today}
\begin{abstract}
We extend the work of Carone, Chaurasia and Vasquez on non-supersymmetric models of flavor based on the 
double tetrahedral group.   Three issues are addressed:  (1) the sector of flavor-symmetry-breaking fields is 
simplified and their potential studied explicitly, (2) a flavorful axion is introduced to solve the strong CP problem 
and (3) the model is extended to include the neutrino sector.  We show how the model can accommodate 
the strong hierarchies manifest in the charged fermion Yukawa matrices, while predicting a qualitatively different form
for the light neutrino mass matrix that is consistent with observed neutrino mass squared differences and mixing angles.
\end{abstract}

\maketitle

\section{Introduction} \label{sec:intro}

The structure of the fermion Yukawa couplings in the standard model may result from the sequential breaking of a horizontal
discrete family symmetry.   Long ago, Aranda, Carone and Lebed~\cite{Aranda:1999kc,Aranda:2000tm} showed how the double tetrahedral 
group $T'$ could be used to construct successful supersymmetric flavor models that are similar to those based on U(2) 
symmetry~\cite{Barbieri:1995uv,Barbieri:1996ww}, with or without the assumption of conventional supersymmetric grand unification.  
For other early work on $T'$ as a flavor symmetry, see Ref.~\cite{early}.  Many other authors have since explored the use of $T'$ symmetry 
in models that aim to address the flavor structure of the standard model~\cite{manytp}.

Much of the work on $T'$ flavor models has assumed weak-scale supersymmetry, to stabilize the hierarchy between the 
weak scale and the grand unified or Planck scale.   Over the past decade, however, there has been no direct evidence for 
superpartners at the LHC, nor indirect evidence in the form of a convincing pattern of deviations from the predictions of
the standard model for some subset of its observables.  While one cannot exclude the possibility that supersymmetry is 
present and just beyond the reach of current experiments (a statement that applies to any new physics that has a decoupling limit), 
the current state of affairs has motivated a greater open-mindedness towards consideration of non-supersymmetric extensions 
of the standard model.  For example, the possibility that the standard model could arise consistently from a string theory without 
supersymmetry has been discussed in Ref.~\cite{Abel:2015oxa}.   The hierarchies between mass scales might result from dynamical 
mechanisms (for example, cosmic relaxation~\cite{Graham:2015cka} or Nnaturalness~\cite{Arkani-Hamed:2016rle}), or anthropic 
selection~\cite{Weinberg:1987dv}.  On the other hand, the 
fundamental mass scales found in nature may simply be random and fine tuned, for reasons that are obscure to us at present.   In this work, we assume the absence of 
supersymmetry and focus on phenomenological issues, while remaining agnostic on the question of naturalness.

The purpose of the present work is to further explore the possibility of nonsupersymmetric models of flavor based on $T'$ symmetry, 
following a study by Carone, Chaurasia and Vasquez~\cite{Carone:2016xsi}.  In Ref.~\cite{Carone:2016xsi}, a nonsupersymmetric $T'$ model 
was presented in which the flavor scale $M_F$ was treated as a free parameter.  (There is less motivation to link the flavor scale to a grand unified scale 
in a framework where the gauge couplings don't automatically unify.)   Global fits were performed to the fermion masses and 
Cabibbo-Kobayashi-Maskawa (CKM) mixing angles, taking into account the nonsupersymmetric running of the Yukawa matrices between the 
scale $M_F$ and the weak scale.  It was found that the model was viable for a wide range of $M_F$;  this scale could be as high as the 
Planck scale or as low as the minimum allowed by the flavor-changing-neutral-current constraints on the heavy, flavor-sector particles with 
masses of order $M_F$.   At the lower end of this range, flavor-sector fields, such as the physical components of the flavon fields that 
spontaneously break the $T'$ symmetry, can potentially have observable consequences.

Here we go beyond the work of Ref.~\cite{Carone:2016xsi} in a number of ways: ({\em i}\hspace{0.1em}) we present a simplification of the 
model involving a smaller number of flavor-symmetry-breaking fields.   While simplicity may be desirable by itself, the smaller field content 
allows a less cumbersome study of the flavon potential that leads to the spontaneous breaking of the flavor symmetry, so that we can confirm 
the assumed pattern of symmetry breaking and study the spectrum of scalar states. ({\em ii}\hspace{0.1em}) We address the strong CP 
problem by promoting an Abelian factor that is required in the model from a $Z_3$ symmetry to an anomalous U(1) symmetry.   This leads to 
a flavorful axion~\cite{Bjorkeroth:2018dzu} (also called a flaxion~\cite{Ema:2016ops}, or 
axi-flavon~\cite{Calibbi:2016hwq,Linster:2018avp,Arias-Aragon:2017eww}, in the literature), 
which leads to more stringent lower bounds on the flavor scale $M_F$ than in our previous study (as well as new avenues for discovery).   
The possibility of flavored axions due to a continuous Abelian factor in a $T'$ flavor model was considered in a supersymmetric model in 
Ref.~\cite{Ahn:2018cau}; the present work gives a simple, nonsupersymmetric realization of this possibility.  ({\em iii}\hspace{0.1em}) We 
extend the model to include the neutrino sector.  As we describe later, one model building difficulty that we must 
overcome is to explain how the small symmetry-breaking parameters that lead to pronounced hierarchies in the charged fermion 
Yukawa matrices lead to much less pronounced hierarchies in the neutrino mass matrix (as indicated, for example, by the two large 
mixing angles).  Our model will show how this outcome can be achieved.

Our paper is organized as follows: in Sec.~\ref{quarks} we present the model and establish our notation.  We study the flavon potential 
including the vacuum alignment and the spectrum of scalar states. We also present a global fit of the charged fermion masses and mixing 
angles, analogous to the one presented in Ref.~~\cite{Carone:2016xsi}.  We address the strong CP problem in Sec.~\ref{axion} and identify the flavored axion couplings to SM particles. Bounds on the axion decay constant from flavor changing decays are given.  
In Sec.~\ref{neutrinos} we address the neutrino sector and introduce a type-I see-saw mechanism with three right-handed neutrinos. In Sec.~\ref{conclusions}, we summarize our
conclusions. 

\section{The Model} \label{quarks}

We assume the flavor symmetry $G_F = T' \times Z_3 \times U(1)$, where the last factor is anomalous and will allow for the existence of a flavorful axion.  We
do not review the group theory of $T'$, which was discussed in some detail in Ref.~\cite{Aranda:2000tm} (including a useful appendix on Clebsch-Gordan factors), and 
reviewed again in Ref.~\cite{Carone:2016xsi}.  We refer the reader to those references for details.   The flavor-symmetry-breaking sector consists of three complex scalar fields 
$A$, $s$, and $\phi$, in the ${\bf 1}^{0-}$, ${\bf 1}^{00}$, and ${\bf 2}^{0+}$ representations of $T'\times Z_3$, using the notation of Ref.~\cite{Aranda:2000tm}.  Notably, the 
triplet flavon $S$ of Ref.~\cite{Carone:2016xsi} has been omitted; the model is nonetheless viable, as we will discuss below.   The complete field content and charge 
assignments for the model are shown in Table~ \ref{charges}.

\begin{table}[h] 
\caption {Charge assignments. The index $a=1,2$ is a generation label. The first four columns correspond to complex scalar fields,
while the remainder are either right-handed standard model fermion fields or Dirac adjoints of left-handed ones.  } \label{charges}
\begin{center}
\begin{tabular}{| c ||  b{0.7cm} |  b{0.7cm} |  p{0.7cm} |  b{0.7cm} ||  b{0.7cm} |  b{0.7cm} |  b{0.7cm} | b{0.7cm} |  b{0.7 cm} | b{0.7 cm} | b{0.7 cm} | b{0.7 cm} | b{0.7cm} | b{0.7cm} | }
 \hline
                              & A                 &$s$          & $\phi$  & $H$      &  $\overline{Q}_L^a$   &  $\overline{Q}_L^3$ &  $d_R^a$  &  $d_R^3$  &  $u_R^a$  &  $u_R^3$ & $\overline{L}^a$ & $\overline{L}^3$ &$e_R^a$  &  $e_R^3$  \\[7pt]
\hline
$T'\times Z_3$        & $\bf{1}^{0-}$&$\bf{1}^{00}$& $\bf{2}^{0+}$      &$\bf{1}^{00}$    &  $\bf{2}^{0-}$ &  $\bf{1}^{00}$ &  $\bf{2}^{0-}$ &  $\bf{1}^{00}$     &  $\bf{2}^{0-}$ &  $\bf{1}^{00}$ & $\bf{2}^{0-}$  & $\bf{1}^{00}$  &$\bf{2}^{0-}$  &  $\bf{1}^{00}$  \\[7pt]
\hline
$U(1)$  & $0$             &   $1$            &$0$      &  $0$       &  $0$              &  $0$               &  $0$          & $-1$        &  $0$ &  $0$    & $0$ & $-1$& $0$ & $0$ \\[7pt]
\hline   
\end{tabular}
\end{center}
\end{table}
Since the standard model fermions are charged under $G_F$, the Yukawa couplings, aside from that of the top quark, arise via 
higher-dimension operators involving the flavon fields. These are suppressed by appropriate powers of the flavor scale $M_F$, the cut 
off of the low-energy effective theory.  When the flavon fields acquire vevs, these operators depend on the ratios 
\begin{equation}
\braket{\phi}/M_F \equiv \left[\begin{array}{c} \epsilon \\ 0 \end{array} \right]\, ,\,\,\,\,\, \braket{A} / M_F \equiv \epsilon' \, , \,\,\,\, \mbox{ and  }\,\,\,\, \braket{s}/M_F \equiv
 \rho \,\,\, .  \label{eq:vevpat}
 \end{equation}
After flavor-symmetry breaking, the following Yukawa textures are generated:  
 \begin{equation}
Y_{U} \sim  \begin{pmatrix}
0      &     u_1  \epsilon'& 0 \\
-u_1  \epsilon'  &   u_2  \epsilon^2   & u_3 \epsilon \\ 
0 & u_4 \epsilon  & u_5
\end{pmatrix} , \label{structureU}
\end{equation}
\begin{equation}
Y_{D} \sim \begin{pmatrix}
0      &     d_1   \epsilon'  & 0 \\
-d_1  \epsilon'  &   d_2  \epsilon^2   & d_3 \epsilon \, \rho \\ 
0 & d_4 \epsilon  & d_5 \rho
\end{pmatrix},  \label{structureD}
\end{equation}  
\begin{equation}
Y_{E} \sim \begin{pmatrix}
0      &     l_1   \epsilon'  & 0 \\
-l_1  \epsilon'  &    l_2  \epsilon^2   & l_3  \epsilon  \\ 
0 & l_4  \epsilon\,  \rho & l_5  \rho
\end{pmatrix}. \label{structureE}  
\end{equation}
Here the $u_i$, $d_i$ and $l_i$ are (in general complex) $\mathcal{O}(1)$ parameters and only the 
leading-order expressions are presented.  The non-zero entries differ in two ways from the textures of 
Ref.~\cite{Carone:2016xsi}: the $2$-$2$ entries above are ${\cal O} (\epsilon^2)$, rather than ${\cal O} (\epsilon)$, 
due to the absence of the $T'$-triplet flavon.  However, the factors of $\rho$ appear in different locations, so that
the end results are qualitatively similar.  For example, the suppression of the $1$-$2$ block of $Y_u$ in Ref.~\cite{Carone:2016xsi} by 
an overall factor of $\rho$ is mimicked here by the higher-order $2$-$2$ entry and the proportionally smaller
numerical value of $\epsilon'$, as we will see later. We also note that there will be CP violation in the 
model even if all the operator coefficients defined at the level of the Lagrangian are real, due to imaginary 
numbers in Clebsch-Gordan coefficients; these would lead, for example, to factors of $i$ in the $2$-$2$ entries of 
$Y_U$, $Y_D$ and $Y_E$.  In general, however, all operator coefficients are themselves complex, and the $10$ phase 
degrees of freedom in $Y_U$ and $Y_D$ can be used to obtain the desired CKM phase rather easily.  In light of this, and to simplify our subsequent
numerical analysis, we have chosen all the operator phases so that the parameters shown in Eqs.~(\ref{structureU})-(\ref{structureE}) are 
real, and omit the CKM phase from our global fit in Sec.~\ref{subsec:fit}.  

 \subsection{The Flavon Potential}  
 
In this subsection, we consider the flavon potential, to confirm that the pattern of vevs assumed in Eq.~(\ref{eq:vevpat}) can be achieved
and to study the spectrum of physical scalar states. We will do this by assuming the desired vev pattern, and imposing the extremization 
conditions on the potential to fix some of its otherwise free parameters.  We then check the second-derivative matrix of the potential for 
positive definiteness.  To simplify the discussion, we exclude the $s$ field, since it is a trivial singlet under the non-Abelian discrete flavor 
group and it is straightforward to write down a potential involving $s$ alone that provides for its vev.  Including terms that couple $s$ to the 
other fields, {\em e.g.}, $|s|^2 |\phi^2|$, will not qualitatively change our results providing that their couplings are not too large, which is good 
enough for a proof of principle.  We are particularly interested in accidental global symmetries that arise in the potential as a consequence of 
the $T' \times Z_3$ discrete symmetry.   These lead to pseudo-goldstone bosons whose masses arise via higher-dimension operators.  We 
estimate the masses of these states to confirm that they are not so light that their phenomenological consequences need to be taken into 
account.  In this case, the only light state that will have interesting flavor-changing physics will be a single flavorful axion associated with the 
$s$ field.
  
 The most general scalar potential for a singlet and a doublet transforming as $ A \sim \bf{1}^{0-}$, $ \phi \sim \bf{2}^{0+}$ under $T' \times Z_3$, respectively, is given by
\begin{equation}
V =  V_{A} + V_{\phi} + V_{A \phi} ,   \label{renpotential} 
\end{equation}
where 
\begin{align}
&V_A =  m_A^2 |A|^2 +  \mu \, (A^3 +  A^{*3}) + \lambda_A \, |A|^4,  \label{VA}\\
&V_{\phi} =  m_\phi^2\,  |\phi|^2 + \lambda_\phi \, |\phi|^4, \label{Vphi} \\
&V_{A \phi} =     \lambda_{A\phi} \, |A|^2 |\phi|^2   \label{VAphi}.  
\end{align}
Note that this potential has an accidental $U(2)_{\phi}$ global symmetry as well as an additional $U(1)_A$ symmetry in the limit $\mu\rightarrow 0$.   We parametrize the fields in terms of their real degrees of freedom
\begin{equation}
A = \frac{1}{\sqrt{2}} (A_1 + i A_2) \,\,\, ,
\label{eq:Adecomp}
\end{equation}
and
\begin{equation}
\phi =\frac{1}{\sqrt{2}} \begin{pmatrix}
             \phi_{11} + i \phi_{12} \\
              \phi_{21} + i \phi_{22}
          \end{pmatrix}.
          \label{eq:phidecomp}
\end{equation}
The Yukawa textures in Eqs.~(\ref{structureU})-(\ref{structureE}) are reproduced provided only the following real fields develop vevs:  
\begin{equation}
\braket{\phi_{11}}/\sqrt{2}=\epsilon \, M_F\,\,\,\,\, \mbox{ and } \braket{A_1}/\sqrt{2} = \epsilon' \, M_F \,\,\, .
\label{eq:vevpat2}
\end{equation}
The location of a local minimum of the potential is determined by six first-derivative equations, corresponding to the six real scalar fields
in Eqs.~(\ref{eq:Adecomp}) and (\ref{eq:phidecomp}).   However, for the assumed vev pattern, only two of these equations are non-vanishing,
\begin{equation}
      \frac{\partial V}{\partial A_1} \biggr\rvert_{\text{vev}} = \sqrt{2} \epsilon' M_F (m_A^2 + 2 \epsilon'^2M_F^2 \lambda_A + \epsilon^2M_F^2 \lambda_{A\phi} + 3 \epsilon' M_F \mu)=0 \,\,\, , 
      \label{eq:fd1}
\end{equation}
\begin{equation}
      \frac{\partial V}{\partial \phi_{11}} \biggr\rvert_{\text{vev}} = \sqrt{2} \epsilon M_F (m_\phi^2 +  \epsilon'^2 M_F^2 \lambda_{A\phi} + 2 \epsilon^2 M_F^2 \lambda_{\phi} )=0 \,\,\,  ,
      \label{eq:fd2}
\end{equation}
were the subscript ``vev" indicates that the fields have been set to their vevs, those shown in Eq~(\ref{eq:vevpat2}) with all others vanishing.  For a given choice of the 
dimensionless couplings, Eqs.~(\ref{eq:fd1}) and (\ref{eq:fd2}) allow us to determine the mass parameters
\begin{equation}
m_A^2 =- 2 \epsilon'^2 M_F^2 \lambda_A - \epsilon^2 M_F^2 \lambda_{A\phi} - 3 \epsilon' M_F \mu \,\,\, ,\label{masstermA}
\end{equation} 
\begin{equation}
m_\phi^2 =- \epsilon'^2 M_F^2 \lambda_{A\phi} - 2 \epsilon^2 M_F^2 \lambda_{\phi}. \label{masstermphi}
\end{equation}
To obtain the mass spectrum, we construct the second derivative matrix for the potential in terms of the six real scalar fields, evaluated
with the assumed vevs, and with mass parameters fixed by Eqs.~(\ref{masstermA}) and (\ref{masstermphi}). In the basis
$(\phi_{11}, A_1, A_2, \phi_{12}, \phi_{21}, \phi_{22})$ we find 
\begin{equation}
m^2_{scalar} = 
\left(\begin{array}{ccclll}  
4\epsilon^2 M_F^2\lambda_\phi  & 2\epsilon \epsilon' M_F^2\lambda_{A\phi} & 0 & 0 \hspace*{1em} & 0 \hspace*{1em} & 0 \hspace*{1em}\\
2 \epsilon \epsilon' M_F^2 \lambda_{A\phi} &  \epsilon' M_F(4 \epsilon' M_F \lambda_A + 3 \mu ) & 0 & 0 & 0 & 0   \\
0 & 0 & -9 \epsilon' M_F \mu & 0 & 0 & 0 \\
0 & 0 & 0 & 0 & 0&0 \\
0& 0 & 0 & 0 & 0 & 0  \\
0 & 0 & 0 & 0 & 0 & 0 
\end{array} \right)
\label{eq:msqmat} 
\end{equation}
The three non-vanishing eigenvalues of the mass squared matrix are positive, provided that $\mu <0$ and (assuming
$\epsilon$, $\epsilon'$ and $\lambda_\phi$ are positive) 
\begin{equation}
|\mu| < \frac{\epsilon' M_F}{3 \lambda_{\phi}} ( 4 \lambda_A \lambda_{\phi} - \lambda^2_{A\phi}) \,\,\, 
\mbox{ and } \,\,\, |\mu| < \frac{4}{3} \frac{M_F}{\epsilon'}(\epsilon^2 \lambda_\phi + \epsilon'^2 \lambda_A) \,\,\, ,
\end{equation}
which is easily arranged. The three massless states are expected from Goldstone's Theorem, since the U(2)$_{\phi}$ symmetry is 
spontaneously broken to a residual U(1) symmetry that rotates the second component of the $\phi$ column vector by a phase.   However, 
these zero eigenvalues are lifted when one takes into account corrections to the potential from higher-dimension operators that break the 
accidental U(2)$_{\phi}$ global symmetry.  We find that the lowest-order operators that have this effect occur at dimension $6$,\begin{equation}
V_{hd} = \frac{1}{M_F^2} \left(c_{61}\left[ (\phi\, \phi)_{\bf{3}} (\phi\, \phi)_{\bf{3}}  \right]_{\bf{3}}  (\phi \,\phi)_{\bf{3}} + \mbox{h.c.}\right)
+ c_{62} \frac{1}{M_F^2} \left[ (\phi^* \,\phi^*)_{\bf{3}} (\phi\, \phi)_{\bf{3}}  \right]_{\bf{3}}  (\phi^*\, \phi)_{\bf{3}} + \cdots \,\, ,
\end{equation}
where the subscript indicates the $T'$ representation of the given product, with Clebsch-Gordan factors left implicit.  We have studied the
eigenvalues of Eq.~(\ref{eq:msqmat}) numerically after including the additional potential terms in $V_{hd}$, and find that all the eigenvalues
are positive and non-vanishing; the masses of the three pseudo-goldstone bosons are of order $\epsilon^2 M_F$.  We will see later that the 
numerical values of our symmetry breaking parameters and our extension to the neutrino sector will imply that this scale corresponds to 
roughly $10^{12}$~GeV.  We therefore do not expect meaningful phenomenological bounds on the three pseudo-goldstone states.  We note 
that there are also dimension-$5$ operators that one can write down which correct the potential  ({\em e.g.}, $A^3 |\phi|^2$ and $A^3 |A|^2$) 
but these do not break the accidental U(2)$_\phi$ symmetry and provide higher-order corrections to the eigenvalues that are already 
non-vanishing at lowest order.

\subsection{Fit to quarks and charged leptons} \label{subsec:fit}

In this subsection, we verify that the Yukawa textures in Eqs.~(\ref{structureU})-(\ref{structureE}) reproduce the correct masses and mixing 
angles for the charged fermions, by performing a global fit that takes into account running from a high scale (which we will take to be 
$4 \times 10^{16}$~GeV, to be consistent with our later discussion of the neutrino sector) down to the weak scale.  This is the same 
analysis that was performed in Ref.~\cite{Carone:2016xsi} for an arbitrary $M_F$ scale, but is now modified to take into account the textures 
predicted in the present model.  We take the model parameters $\{ u_i,d_i, l_i, \epsilon, \epsilon',\rho \}$ to be real as a simplifying 
assumption since, as discussed earlier, there is no difficulty in accommodating a CKM phase if one allows an arbitrary phase 
parameter for every operator coefficient.  The experimental inputs are the quark and lepton masses and CKM angles, which we associate 
with the scale $m_Z$  ({\em i.e.}, we ignore weak scale threshold corrections).   We seek solutions in which the order one parameters are in fact 
not far from one,  while predictions for the observables, renormalized at the weak scale, are within two standard deviations of experimental values.  
Employing the same technique as Refs.~\cite{Aranda:2000tm} and \cite{Carone:2016xsi}, we construct a function $\widetilde{\chi}$ whose minimization 
achieves this goal:
\begin{equation}
\begin{aligned}
\widetilde{\chi}^2 &= \sum_{i=1}^9 \left(\frac{m_i^{th}-m_i^{exp}}{\Delta m_i^{exp}}\right)^2+\left(\frac{|V_{us}^{th}|-|V_{us}^{exp}|}{\Delta V_{us}^{exp}}\right)^2+\left(\frac{|V_{ub}^{th}|-|V_{ub}^{exp}|}{\Delta V_{ub}^{exp}}\right)^2+\left(\frac{|V_{cb}^{th}|-|V_{cb}^{exp}|}{\Delta V_{cb}^{exp}}\right)^2
\\ &+\sum_{i=1}^5 \left(\frac{\ln |u_i|}{\ln 3}\right)^2+\sum_{i=1}^5 \left(\frac{\ln|d_i|}{\ln 3}\right)^2+\sum_{i=1}^5 \left(\frac{\ln|\ell_i|}{\ln 3}\right)^2 
\,\,\, .
\end{aligned} \label{chifun}
\end{equation}
The first four terms would be present in a conventional chi-squared function, and place weight on how close the theoretical predictions for
observables are to experimental observations, relative to the experimental error. The experimental errors are handled as in  
Refs.~\cite{Aranda:2000tm, Carone:2016xsi}: they are inflated to $1\%$ of the central measured value if the error is smaller that this amount.  
This takes into account theoretical uncertainties (for example, two-loop running effects) that have been omitted.  The remaining three terms 
of Eq.~(\ref{chifun}) place weight on the coefficients having values that are order one, the expectation of naive dimensional analysis.   
Including these terms is equivalent to assuming that the coefficients are distributed with a log-normal distribution with mean 1 and 
standard deviation $\sigma = \ln{(3)}/2$ such that the absolute value of an element drawn from the distribution has a $95\%$ probability 
to lie in the range $[0.3, 3]$. There are a total of $12$ observables (nine masses and three mixing angles) and, given the stated 
constraints on the model parameters, the only three genuine free parameters, $\{ \epsilon, \epsilon',\rho \}$.  Thus, we expect a good 
fit if $\widetilde{\chi}^2 \approx 9$. The best fit values together with the experimental and theoretical predictions are presented in 
Table~\ref{tableQuarks}.  We note that our successful results might be anticipated from the qualitatively similar Yukawa textures obtained
in U(2) flavor models~\cite{Linster:2018avp,Falkowski:2015zwa}, a further example of the similarities between $T'$ models and U(2) models 
that was the focus of Refs.~\cite{Aranda:1999kc,Aranda:2000tm}.

Finally, we note that global symmetries are expected to be broken by quantum gravity effects~\cite{Krauss:1988zc}, but we can assume that
there is an ultraviolet completion which allows the U(1) symmetry to arise as a consequence of the continuous and discrete gauge 
symmetries that are present in a more complete theory.   Another concern in the present framework is that the breaking of discrete 
symmetries can lead to potential domain-wall problems.  However, these can be rendered harmless it the domain regions are widely 
separated due to inflation.   We will find later in Sec.~\ref{axion} that $M_F$ is constrained to be sufficiently high so that any problems with 
domain walls may be eliminated via this mechanism.

\begin{center} 
\begin{table}[h]
\caption{Fit to the charged fermion masses and mixing angles.  All masses are given in GeV. (Note that $m_t$ is 
the $\overline{MS}$ mass, not the pole mass.)  The value of the quantity $\widetilde{\chi}^2$ defined in the text 
is $12.3$.  Running from the flavor scale $M_F$ down to the $Z$ mass is taken into account, with 
$M_F = 4 \times 10^{16}$~GeV, (see Sec.~\ref{neutrinos}) chosen for the purpose of illustration.}
\label{tableQuarks}
\vspace{2mm}
\label{tbl:res1}
\begin{tabular}{ccc} \hline\hline
\multicolumn{1}{c}{} &
\multicolumn{1}{c}{Best Fit Parameters} &
\multicolumn{1}{c}{}\\
\hline
\multicolumn{1}{c}{} &
\multicolumn{1}{c}{$\epsilon=2.42\times 10^{-2}, \; \epsilon'=9.75 \times 10^{-5},\; \rho=-1.38\times 10^{-2} $} &
\multicolumn{1}{c}{}\\
\hline
$u_{1}= 1.22$  & $d_{1}=0.662$  & $\ell_{1}= 0.612$ \\
$u_{2}=-0.671$  &  $d_{2}=1.29$ & $\ell_{2}=0.643$ \\
$u_{3}=-2.26$ & $d_{3}=-1.02$  & $\ell_{3}=0.352$ \\
$u_{4}= -0.702$   & $d_{4}=-0.276$  & $\ell_{4}=2.40$ \\
$u_{5}=0.384$ & $d_{5}=0.376$ & $\ell_{5}=0.295$  \\
\hline \hline
\multicolumn{1}{c}{Observable}&
\multicolumn{1}{c}{Expt. Value from \cite{Tanabashi:2018oca}}&
\multicolumn{1}{c}{Fit Value}\\
\hline
$m_{u}$    & $(2.2 \pm 0.45)\times 10^{-3}$       & $2.30 \times 10^{-3}$ \\
$m_{c}$    & $1.275   \pm 0.03$        & 1.274 \\
$m_{t}$    & $160      \pm 4.5$          & 160.0  \\
$m_{d}$    & $(4.7   \pm 0.4)\times 10^{-3}$       & $5.42\times 10^{-3}$ \\
$m_{s}$    & $(9.5 \pm 0.6)\times 10^{-2}$        & $9.16 \times 10^{-2}$ \\ 
$m_{b}$    & $4.18     \pm 0.035$         & 4.17 \\
$m_{e}$    & $(5.11 \pm 1\%)\times 10^{-4}$ & $5.11\times10^{-4}$ \\ 
$m_{\mu}$  & $0.106    \pm 1\%$      & 0.106 \\
$m_{\tau}$ & $1.78     \pm 1\%$       & 1.78\\
$|V_{us}|$ & $0.225 \pm 1\%$    & 0.223 \\ 
$|V_{ub}|$ & $(3.65 \pm 0.12)\times10^{-3}$      & $3.62\times10^{-3}$ \\ 
$|V_{cb}|$ & $(4.21 \pm 0.08)\times10^{-2}$       & $4.17 \times 10^{-2}$ \\
\hline\hline
\end{tabular}
\end{table}
\end{center}
  
\newpage
 \section{The Flavorful Axion} \label{axion}

The model we have presented includes a flavon field $s$, charged under the U(1) factor of the flavor group, which assures, for example, 
the correct values of the bottom quark and tau lepton Yukawa couplings.  This U(1) also serves as a Peccei-Quinn (PQ) symmetry and its 
spontaneous breaking leads to a flavorful axion.  Only the third generation right-handed down quark and the third generation left-handed 
lepton doublet have nontrivial charges under the U(1) symmetry (see Table~\ref{charges}), but rotation to the mass eigenstate basis will 
induce axion couplings to fields of the first two generations.   The axion is identified via the non-linear representation
 \begin{equation}
 s =  \frac{v_s + \sigma}{\sqrt{2}} e^{i a/v_s} \,\,\, .
 \end{equation} 
The radial component  $\sigma $ is a heavy degree of freedom and is integrated out of the low-energy effective field theory. The phase 
field $a$ is the Goldstone boson of the spontaneously broken U(1)$\equiv$U(1)$_{PQ}$ symmetry and is identified with the QCD axion.  
Non-perturbative QCD effects generate a potential for the axion, with the minimum corresponding to vanishing of the $\overline{\theta}$ 
parameter of QCD, solving the strong CP problem.  For complete reviews on this subject see Refs.~\cite{Kim:1986ax,Marsh:2015xka}.    
 
The axiflavon couplings to fermions originate from the following Yukawa couplings   
\begin{equation}
 \mathcal{L}_{Y a} = -\left[ \overline{Q}^{\,i} Y_{i3}^d  H d_R^3 + \overline{L}^3 Y_{3j}^e H e_R^j \right]\frac{s}{M_F} + \mbox{ h.c.}\, ,
\end{equation} 
or more explicitly
\begin{eqnarray} 
 \mathcal{L}_{Y a} &=& -\left[ d_3 \overline{Q}^a \epsilon_{ab} \left( \frac{ \langle \phi^b \rangle}{M_F}\right) H d_R^3 +  d_5 \overline{Q}^3 H d_R^3 \right. 
 \nonumber \\
 &+& \left. l_4  \overline{L}^3 \epsilon_{ab}  \left( \frac{\langle \phi^b \rangle}{M_F}\right)  H e_R^a + l_5  \overline{L}^3 H e_R^3 \right] \frac{v_s \, e^{ia/v_s}}{\sqrt{2}M_F} + \mbox{ h.c.} \, ,    
\end{eqnarray}
where $\epsilon_{ab}$, $a,b=1,2$ is the Clebsch-Gordan matrix that allows one to combine two ${\bf 2}^0$ representations of $T'$ into a ${\bf 1}^0$.  Setting the $\phi$ flavon
to its vev, one obtains
\begin{equation}  
\mathcal{L}_{Y a}  =  -\left[- d_3 \epsilon \, \overline{Q}^2  H d_R^3 +  d_5 \overline{Q}^3 H d_R^3 - l_4  \epsilon \, \overline{L}^3 H e_R^2 + l_5  \overline{L}^3 H e_R^3 \right]\frac{v_s\, e^{ia/v_s}}{\sqrt{2}M_F}+ \mbox{ h.c.}
\end{equation}
Performing the usual non-linear field redefinition of the third generation fermions
  \begin{equation}
 d_R^3  \rightarrow e^{-i a/v_s} d_R^3,   \quad \overline{L}^3  \rightarrow  e^{-i a/v_s} \overline{L}^3,   \label{axionrotation}
 \end{equation}
we remove the axion entirely from the Yukawa sector, but instead induce derivative interactions coming from the original fermion kinetic terms.  For the charged fermions, one finds
\begin{equation}
\mathcal{L} \supseteq \frac{\partial_\mu a}{v_s}  \left[ \bar{d}_i \gamma^\mu  (K_d^\dagger)_{i3}(K_d)_{3j}   \frac{(1+\gamma_5)}{2} d_j  +\bar{e}_i \gamma^\mu  (U_e^\dagger)_{i3}(U_e)_{3j}   \frac{(1- \gamma_5)}{2} e_j\right]  \,\,\, .\label{axionFCNC}
\end{equation}
Here $K_d$ ($U_e$) is the right-handed (left-handed) rotation that diagonalize the Yukawa interactions, where in our conventions a
generic Yukawa matrix $Y$ would be diagonalized by $Y = U_L Y^{diag} U_R^\dagger$.  Notice that the axion interactions with the fermion 
mass eigenstates are in general not diagonal and therefore induce flavor-changing neutral currents (FCNC) at tree-level. Flavon FCNCs are very well constrained experimentally 
\cite{Feng:1997tn, Bauer:2016rxs} and we will discuss these constraints in the next subsection.  See Refs.~\cite{Celis:2014iua, Albrecht:2010xh, Davidson:1984ik} for other axion 
models with FCNCs at tree-level. 

While our phenomenological bounds will come from the couplings in Eq.~(\ref{axionFCNC}), we give the axion couplings to two gauge fields here for 
completeness.  After the anomalous chiral rotation in Eq.~(\ref{axionrotation}), the axion reappears in an effective interaction with the gluon field strength 
and its dual, namely
\begin{equation}
\mathcal{L} = \frac{\alpha_s}{8\pi}\frac{a}{v_s}N_{DW} G_{\mu \nu}^a\widetilde{G}^{a \mu \nu}.
\end{equation} 
 With the charge assignments of Table~\ref{charges}, we obtain the domain-wall number  
 \begin{equation}
 N_{DW} = \left[2\sum_{i}X_Q^i - \sum_{i}X_u^i-\sum_i X_d^i  \right] = 1  \,\,\, ,
 \end{equation}
where the $X_a$ represent the U(1) charges for left-handed and the right-handed fermion fields.  Since $N_{DW}=1$, there is one minimum of the axion potential.  We identify the axion decay 
constant as
 \begin{equation}
 f_a = |v_s / N_{DW}| \,\,\, .
 \end{equation}
The PQ charge assignments give rise to U(1)$_Y^2$ U(1)$_{PQ}$ and SU(2)$^2 U(1)_{PQ}$ anomalies and therefore axion couplings to hypercharge and electroweak gauge bosons are induced, namely
\begin{equation}
\mathcal{L} \supseteq \frac{g'^2}{32\pi^2}\frac{a}{v_s}(2N_{B}) B_{\mu \nu}\widetilde{B}^{ \mu \nu} +\frac{g^2}{32\pi^2}\frac{a}{v_s}N_{W} W_{\mu \nu}^a\widetilde{W}^{a \mu \nu}. 
\end{equation} 
Rewriting this piece of the Lagrangian in the gauge boson mass eigenstate basis one obtains the axion couplings to photons
\begin{equation}
\mathcal{L}_{\gamma a} = \frac{\alpha_{EM}}{8\pi}\frac{a}{v_s}(2N_{B} + N_W) F_{\mu \nu}\tilde{F}^{ \mu \nu}
\end{equation}
where in this model one obtains
\begin{align}
N_B &= 3 \left[ 2 \sum_i \left(\frac{1}{6} \right)^2 X_Q^i  -  \sum_i \left(\frac{2}{3}\right)^2 X_u^i - \sum_i \left(-\frac{1}{3}\right)^2 X_d^i \right] \\             
         &  + 2 \sum_i \left(- \frac{1}{2} \right)^2 X_L^i -  \sum_i (-1)^2 X_e^i  = \frac{5}{6},
\end{align}
\begin{equation}
N_W =  \sum_i X_L^i + 3 \sum_i X_Q^i = 1,
\end{equation}
and thus the ratio of the electromagnetic to color anomalies is 
\begin{equation}
\frac{2N_B + N_W}{N_{DW}}= \frac{8}{3}.
\end{equation}
As noted in other flavored axion models that make the same prediction for this ratio~\cite{Calibbi:2016hwq}, this is consistent with the predictions of the simplest DFSZ axion models~\cite{Kim:1986ax,Marsh:2015xka}.

\subsection{ Constraints from meson decays}

As can be seen from the axion couplings to fermions in Eq.~\eqref{axionFCNC}, our model predicts flavor violating processes, {\em e.g.}, 
heavy meson decays like $K^+\rightarrow \pi^+ a$.  The branching fraction for a generic meson two-body decay $P\rightarrow P' \, a$ is given by~\cite{Bjorkeroth:2018dzu}
\begin{equation}
BR(P\rightarrow P'a) = \frac{1}{64\pi \Gamma(P)} \frac{|(K_d)^\dagger_{i3} (K_d)_{3j}|^2}{f_a^2} m_P^3 \left( 1-\frac{m_{P'}^2}{m_P^2} \right)^3|f_+(0)|^2
\end{equation}
where $P=(\bar{q}_iq)$, $P'=(\bar{q}_j q)$ and the indices $ij$ denote the constituent quarks. The function $f_+(q^2)$ is the form factor from hadronic physics calculations 
and $q=q_P-q_{P'}$ is the momentum transfer to the axion; one may take $q^2\approx 0$ as the axion is very light.  The axion mass is the same as a QCD axion, $m_a \approx 6 \times 10^{-6}  \cdot (10^{12} \mbox{ GeV}/f_a) \,\,\, \mbox{eV}$~\cite{Bjorkeroth:2018dzu}; we will see that the strongest bounds presented later in this section imply $m_a \alt 10^{-4}$~eV, while the neutrino model discussed in the next section corresponds to $m_a \approx 7 \times 10^{-9}$~eV.

Experimental bounds on different heavy mesons decays are summarized in Ref.~\cite{Bjorkeroth:2018dzu}.  In
Table~\ref{Mesonconstraints}, we quote the most relevant of these constraints and indicate the relevant experimental references.   
The precise numerical bounds that follow from the fit presented in Sec.~\ref{subsec:fit} are displayed in the last column of this table.

To understand our results qualitatively, it is useful to parameterize the rotation matrices that correspond to the fit in Table~\ref{tbl:res1} in terms of powers of the Cabibbo angle $\lambda \approx 0.22$.  We find numerically that $K_d$ and $U_e$ have the qualitative form
 \begin{equation}
K_{d} \sim  \begin{pmatrix}
1      &      \lambda &  \lambda^5 \\
\lambda  &  1  & 1 \\ 
 \lambda & 1  &1
\end{pmatrix}
\,\,\, \mbox{ and } \,\,\,
U_{e} \sim  \begin{pmatrix}
1      &      \lambda &  \lambda^5 \\
\lambda^2  &  1  & 1 \\ 
 \lambda^2 & 1  &1
\end{pmatrix}  \,\,\, .
\end{equation} 
The relevant combinations that determine the results in Tables~\ref{Mesonconstraints} and \ref{Leptonconstraints} are
\begin{equation}
 (K_d^\dagger)_{i3}(K_d)_{3j} \sim  \begin{pmatrix}
\lambda^2      &      \lambda &  \lambda \\
\lambda  &  1  & 1 \\ 
 \lambda & 1  &1
\end{pmatrix}
\,\,\, \mbox{ and } \,\,\,
 (U_e^\dagger)_{i3}(U_e)_{3j}  \sim
\begin{pmatrix}
\lambda^3      &      \lambda^2 &  \lambda^2 \\
\lambda^2  &  1  & \lambda \\ 
 \lambda^2 & \lambda  &1
\end{pmatrix}.
\end{equation}

\begin{table}[h]
\caption{Experimental constraints on the branching fractions of heavy mesons decays (second column),  derived bounds on the axion decay constant times flavor rotation matrix elements from Ref.~\cite{Bjorkeroth:2018dzu} (third column) and lower bound on the axion decay constant using the numerical value of the matrix element from the fit presented in Sec.~\ref{subsec:fit} (fourth column).}\label{Mesonconstraints}
\begin{tabular}{|l|l|l|l|l}
\cline{1-4}
\multicolumn{1}{|c|}{\text{Decay}} & \multicolumn{1}{c|}{ \text{Branching Ratio} }             & \multicolumn{1}{c|}{\text{Bound ($f_a/\GeV$)}}                   &    \text{Bound from fit} &  \\ \cline{1-4}
  \textit{$K^+ \rightarrow \pi^+ a$} &   \textit{$<0.73\times10^{-10}$}\cite{Adler:2008zza} &\textit{$>3.45\times 10^{11} |(K_d^\dagger)_{23} (K_d)_{31}|$} & $f_a > 6.3\times10^{10} \mbox{ GeV}$ &  \\ \cline{1-4}
\textit{$K_L^0 \rightarrow \pi^0 a$} & \textit{$<5\times10^{-8}$}   \cite{Ahn:2016kja}    &\textit{$>1.35\times 10^{10} |(K_d^\dagger)_{23} (K_d)_{31}|$} & $f_a > 2.5\times10^{9} 
\mbox{ GeV}$  &  \\ \cline{1-4}
\multicolumn{1}{|l|}{\textit{$B^\pm \rightarrow \pi^\pm a$}} & \multicolumn{1}{l|}{ \textit{$<4.9\times10^{-5}$} \cite{Ammar:2001gi}} &  \textit{$>5.0\times 10^{7} |(K_d^\dagger)_{33} (K_d)_{31}|$}& $f_a > 7.4 \times10^6 \mbox{ GeV}$ & \\ \cline{1-4}
\textit{$B^\pm \rightarrow K^\pm a$}              & \textit{$<4.9\times10^{-5}$} \cite{Ammar:2001gi}          & \textit{$>6.0\times 10^{7} |(K_d^\dagger)_{33} (K_d)_{32}|$}        & $f_a > 2.8\times10^7 \mbox{ GeV}$ & \\ \cline{1-4}
\end{tabular}
\end{table}
The strongest bound in this model comes from the heavy meson decay $K^+ \rightarrow \pi^+ a$ giving
\begin{equation}
f_a > 6.3\times10^{10} \mbox { GeV}.
\label{eq:kbound}
\end{equation}
Given the identification $f_a=|v_s/N_{DW}| = \sqrt{2} |\rho| M_F$, we can translate this to a bound on the flavor scale
\begin{equation}
M_F > 3.2\times 10^{12} \mbox{ GeV}. \label{mesonbound}
\end{equation}
Axion mixing with neutral hadronic mesons does not lead to competitive bounds and will not be discussed here. See Ref.~\cite{Bjorkeroth:2018dzu} for a treatment of 
these effects. 

\subsection{ Constraints from lepton decays}

\begin{table}[h]
\caption{Experimental constraints on the branching fractions of lepton decays (second column), derived bounds on the axion decay constant times flavor rotation matrix elements from Ref.~\cite{Bjorkeroth:2018dzu}  (third column) and lower bound on the axion decay constant using the predicted numerical values from out fit (fourth column).}\label{Leptonconstraints}
\begin{tabular}{|c|c|c|l|l}
\cline{1-4}
\text{Decay}              & \text{Branching Ratio}             & \text{Bound ($f_a/\GeV$)}             & \text{Bound from fit}  &  \\ \cline{1-4}
\multicolumn{1}{|l|}{\textit{$\mu^+ \rightarrow e^+ a$}  } & \multicolumn{1}{l|}{\textit{$<1.0 \times10^{-5}$}\cite{Bayes:2014lxz}} & \multicolumn{1}{l|}{ \textit{$>2.0 \times 10^{9} |(U_e^\dagger)_{23} (U_e)_{31}|$}} & $f_a>1.7\times10^8\mbox { GeV}$ &  \\ \cline{1-4}
\textit{$\tau^+ \rightarrow e^+ a$}     & \textit{$<1.5\times10^{-2}$}   \cite{Albrecht:1995ht}    & \textit{$>1.3\times 10^{6} |(U_e^\dagger)_{33} (U_e)_{31}|$}         & $f_a>5.3 \times 10^4\mbox{ GeV}$ &  \\ \cline{1-4}
\multicolumn{1}{|l|}{\textit{$\tau^+ \rightarrow \mu^+ a$} } & \multicolumn{1}{l|}{ \textit{$<2.6\times10^{-2}$} \cite{Albrecht:1995ht}} & \multicolumn{1}{l|}{\textit{$>9.9\times 10^{5} |(U_e^\dagger)_{33} (U_e)_{32}|$} } & $f_a>3.9 \times 10^5 \mbox{ GeV}$ &  \\ \cline{1-4}
\end{tabular}
\end{table}

From the axiflavon couplings in Eq.~\eqref{axionFCNC} one can also compute the branching fraction for leptonic decays, namely~\cite{Bjorkeroth:2018dzu} 
\begin{equation}
BR(e_i \rightarrow e_j a) =  \frac{1}{32\pi \Gamma(e_i)}\frac{m_i^3}{f^2_a} |(U_e^\dagger)_{i3}(U_e)_{3j}|^2\left( 1- \frac{m_j^2}{m_i^2} \right)^3 \,\, .
\end{equation}
The most stringent bound comes from the decay $\mu^+ \rightarrow e^+a $ giving $f_a > 1.7\times 10^8 \mbox{ GeV}$, which is not competitive with our earlier bound from charged kaon decays, Eq.~(\ref{eq:kbound}).

One can also find bounds from lepton decays with a photon in the final state but it turns out that these are not stronger than the bounds we have already considered.

\section{Neutrino Sector} \label{neutrinos}
In this section, we consider how our model may be extended to explain the observed neutrino masses and mixing angles.  In doing so, we face an immediate challenge: how can we 
explain two large neutrino mixing angles in a theory where symmetry breaking is achieved through two small parameters, $\epsilon$ and $\epsilon'$, that are of 
order $10^{-2}$ and $10^{-4}$, respectively?   A similar problem presents itself when one considers the neutrino mass squared differences.   The smallness of the overall neutrino 
mass scale can be explained via the see-saw mechanism; we will implement a type-I see-saw mechanism below, involving three right-handed neutrinos.  Choice of the right-handed 
neutrino mass scale allows us to fix one of the observed neutrino mass squared differences, for example, $\Delta m^2_{32}$; what is then determined by the symmetry breaking 
parameters is the ratio $\Delta m^2_{32}/\Delta m^2_{21}$, which is found experimentally to be $33.3\pm1.03$~\cite{Tanabashi:2018oca}, 
assuming a normal, rather than inverted, neutrino mass hierarchy (which is the case on our model).   One would expect that the theoretical 
prediction for $\Delta m^2_{32}/\Delta m^2_{21}$ is proportional to ratios of powers of  $\epsilon$ and $\epsilon'$; if this quantity is not 
$\mathcal{O}(1)$, then one finds typically that the predicted value is either much too large or too small  to account for the 
experimental value. This is a consequence of the small and distinctly hierarchical values of $\epsilon$ and $\epsilon'$.   On the other hand, if the ratio $\Delta m^2_{32}/\Delta m^2_{21}$ is approximately independent of $\epsilon$ and $\epsilon'$, then it is a function of the order one coefficients in the theory alone.  In this case, a value of $33.3$ can be obtained for a rather mundane reason:   The see-saw formula tells us that the mass matrix of the light, left-handed neutrino mass eigenstates is given by
\begin{equation}
M_{LL} \approx M_{LR} M_{RR}^{-1} M_{LR}^\dagger \,\,\, ,
\end{equation}
which implies that the eigenvalues of $M_{LL}$ will typically be of cubic order in quantities of ${\cal O}(1)$, either operator coefficients or 
their inverse.  Here, $M_{LR}$ represents the neutrino Dirac mass matrix, while $M_{RR}$ is the Majorana mass matrix for the right-handed 
neutrinos. The numerator and denominator of $\Delta m^2_{32}/\Delta m^2_{21}$ then each depend on terms that are of {\em sixth} order in 
quantities that are ${\cal O}(1)$, with each typically falling somewhere between $1/3$ and $3$ in absolute value, given our earlier 
assumptions.  Noting that $1.8^6 \approx 34$, one can understand how easy it is to take input matrices with coefficients that are of 
${\cal O}(1)$ and still obtain a mass-squared-difference ratio that is consistent with the experimental value.   This observation is relevant to 
our solution below.

We introduce three right-handed neutrinos that are uncharged under the Peccei-Quinn symmetry and have $T' \times Z_3$ charges
\begin{equation}
\nu_R^1 \sim {\bf 1}^{0-}, \,\,\,\,\, \mbox{ and } \,\,\,\,\, \nu_R^{2,3} \sim \bf{1}^{00} \,\,\, .
\end{equation}
The Dirac and Majorana mass matrices have the following $T' \times Z_3 \times U(1)$ transformation properties
\begin{equation}
M_{LR} \sim \left(\begin{array}{cc|c} {\bf 2}^{0-} & {\bf 2}^{0+} & {\bf 2}^{0+} \\ \hline 
{\bf 1}^{0+}_{+1} & {\bf 1}^{00}_{+1} & {\bf 1}^{00}_{+1} \end{array} \right)
\,\,\,\,\, \mbox{ and } \,\,\,\,\,
M_{RR} \sim \left(\begin{array}{cc|c} {\bf 1}^{0-} & {\bf 1}^{0+} & {\bf 1}^{0+} \\ 
{\bf 1}^{0+} & {\bf 1}^{00} & {\bf 1}^{00} \\ \hline 
{\bf 1}^{0+} & {\bf 1}^{00} & {\bf 1}^{00}\end{array} \right) \,\,\, ,
\end{equation}
where we have indicated U(1) charges with a subscript.  This leads to the textures
 \begin{equation}
M_{LR} = \frac{v}{\sqrt{2}}  \begin{pmatrix}
b_1 \epsilon  &     0   &  0 \\
0                   &  b_2 \epsilon   & b_3 \epsilon \\ 
b_4\rho\, \epsilon' & b_5\rho &  b_6 \rho
\end{pmatrix} , \,\,\,\,\, \mbox{ and } \,\,\,\,\,
M_{RR} = \begin{pmatrix}
c_1 \epsilon' M_F     &   c_2    \epsilon' M_F & c_3 \epsilon' M_F \\
c_2  \epsilon' M_F &  M_{22}   & M_{23} \\ 
c_3 \epsilon' M_F &  M_{23}  &  M_{33}
\end{pmatrix}.   \label{structureMRR}
\end{equation}
Here the $b_i$ and $c_i$ are $\mathcal{O}(1)$ coefficients.  Since the elements labelled $M_{22}$, $M_{23}$ and $M_{33}$ in $M_{RR}$ are each flavor-group 
invariant, they don't necessarily have to be at the same scale as $M_F$, or as each other.  For the purposes of demonstrating the viability of the neutrino sector, we will 
take these elements to be at the scale $\epsilon' \, M_F$, so that $M_{RR}$ takes the form
\begin{equation}
M_{RR} = \epsilon' M_F \begin{pmatrix}
c_1   &   c_2     & c_3  \\
c_2   &   c_4     &  c_5 \\ 
c_3   &   c_5     &  c_6
\end{pmatrix} \equiv \epsilon' M_F \widetilde{M}_{RR} .   \label{eq:simpleRR}
\end{equation}
In other words, with this choice, the right-handed Majorana matrix is a complete arbitrary matrix with $\mathcal{O}(1)$ entries, $\widetilde{M}_{RR}$, times the scale $\epsilon' M_F$.   
The Dirac mass matrix also has considerable freedom.   Noting that our earlier fits indicated $\rho \approx {\cal O}(\epsilon)$, we can redefine the coefficients $b_5$ and $b_6$, and
drop the $13$ entry, which is higher order.   Then we see that $M_{LR}$ is approximately of the form
\begin{equation}
M_{LR} \approx \frac{v \, \epsilon}{\sqrt{2}}  \begin{pmatrix}
b_1 &     0   &  0 \\
0                   &  b_2    & b_3  \\ 
0                  & b_5  &  b_6 
\end{pmatrix} \equiv  \frac{v \, \epsilon}{\sqrt{2}} \left(\begin{array}{cc} b_1 & 0 \\
0 & \widetilde{Y}_{LR} \end{array} \right) \,\,\, ,  \label{eq:simpleLR}
\end{equation}
where $\widetilde{Y}_{LR}$ is an arbitrary, two-by-two matrix with $\mathcal{O}(1)$ entries.  The $10$ free parameters in 
Eqs.~(\ref{eq:simpleRR}) and the approximation shown in (\ref{eq:simpleLR}) 
are more than sufficient to obtain the desired values of $\Delta m^2_{32}/\Delta m^2_{21}$, as well as $\sin^2\theta_{12}$, 
$\sin^2\theta_{13}$ and $\sin^2\theta_{23}$, while maintaining $\mathcal{O}(1)$ operator coefficients.  The dependence of the output on 
products of the coefficients allows numerical values like $33$ (the experimental value of 
$\Delta m^2_{32}/\Delta m^2_{21}$) or $1/33$ (very close to $\theta_{13}^2$) to arise without fine tunings.  We note that the form of 
Eq.~(\ref{eq:simpleLR}), with a non-vanishing $1$-$1$ entry, is a consequence of the different charge assignment for the first-generation 
right-handed neutrino field.  This entry of $M_{LR}$ originates from 
a charge conjugated ${\bf 2}^{0+}$ flavon; in $T'$, as in SU(2), ${\bf 2} \sim i \sigma^2 {\bf 2}^*$, which flips the relative location of the
doublet vev in the first two columns of $M_{LR}$.

\begin{table}[h]
\caption{Example of a viable parameter choice for the neutrino sector. \label{tableNeutrinos}}
\vspace{2mm}
\begin{tabular}{ccc} \hline\hline
\multicolumn{1}{c}{} &
\multicolumn{1}{c}{Parameters} &
\multicolumn{1}{c}{}\\
\hline
\multicolumn{1}{c}{} &
\multicolumn{1}{c}{$\epsilon=2.42\times 10^{-2}, \; \epsilon'=9.75 \times 10^{-5},\; \rho=-1.38\times 10^{-2} $} &
\multicolumn{1}{c}{}\\
\hline
$b_{1}= 1.66$  &   $b_{2}=  1.07$ &  $b_{3}=  2.10$   \\
$b_{4}=1.11$  &  $b_{5}=  -0.891$  & $b_{6}= 1.61$   \\
$c_{1}= 2.91$ & $c_{2}=    1.04$ & $c_{3}= 0.662$\\
$c_{4}= 1.21$ & $c_{5}=    1.37$ & $c_{6}=    1.44$  \\
\hline \hline
\multicolumn{1}{c}{Observable}&
\multicolumn{1}{c}{Expt. Value from \cite{Tanabashi:2018oca}}&
\multicolumn{1}{c}{Fit Value}\\
\hline
$\frac{\triangle m_{32}^2}{\triangle m_{21}^2}$  & $33.3 \pm 1.03$                          & $33.8$ \\
$\sin^2{\theta_{12}}$                                       & $ 0.307  \pm 0.013$                      &  0.307 \\
$\sin^2{\theta_{23}}$                                       & $0.417     \pm 0.025$                    & 0.444  \\
$\sin^2{\theta_{13}}$                                      & $(2.12   \pm 0.08)\times 10^{-2}$    & $2.11\times 10^{-2}$ \\
\hline\hline
\end{tabular}
\end{table}

An example of a viable parameter set for the neutrino sector is shown in Table~\ref{tableNeutrinos}.   The neutrino mixing angles are defined via a 
standard parametrization of the PMNS matrix, which we call $U$ below, 
\begin{equation}
U = U_e^\dagger U_\nu,
\end{equation}
where $U_e$ ($U_\nu$) is a unitary matrix that diagonalizes the charged lepton (left-handed Majorana) matrix following our earlier
convention, {\em i.e.}, $M_{LL} = U_\nu M_{LL}^{diag} U_\nu^\dagger$.  
We can extract the mixing angles via the relations
\begin{equation}
\sin^2\theta_{13} = U_{13}^2 \,\,\,\,, \,\,\,\, \sin^2\theta_{23} = U_{23}^2/(1-U_{13}^2) \,\,\,\, \mbox{ and } \,\,\,\,  \sin^2\theta_{12} = U_{12}^2/(1-U_{13}^2) \,\,\, .
\end{equation}
For the purpose of illustration, we fix  $\epsilon$, $\epsilon'$ and $\rho$, as well as the coefficients $l_i$ appearing in the charged lepton 
Yukawa matrix, to the values that were obtained in our previous global fit of the charged fermions, Table~\ref{tbl:res1}.  A viable choice of 
neutrino sector parameters $b_i$ and $c_i$ is presented in Table~\ref{tableNeutrinos}.  These
were obtained by defining a $\widetilde{\chi}_\nu^2$ for the neutrino sector that takes into account the neutrino observables listed in the 
table and also places weight on the neutrino-sector coefficients being $\mathcal{O}(1)$, in analogy to our approach in the charged fermions.  
This function can be used to diagnose when a good-enough parameter choice has been obtained.

Since the right-handed neutrino mass scale is set by $\epsilon' M_F$, the neutrino mass squared differences (rather than the ratio) can be used to determine the 
flavor scale.  Using either experimental value~\cite{Tanabashi:2018oca}
\begin{equation}
\triangle m_{21}^2 = (7.53 \pm 0.18)\times 10^{-5} \eV^2 \,\,\,\, \mbox{ or }\,\,\,\, \triangle m_{32}^2 = (2.51 \pm 0.05)\times 10^{-3} \eV^2,
\end{equation}
we find that the solution in Table~\ref{tableNeutrinos} corresponds to 
\begin{equation}
M_F = 4.6 \times 10^{16} \mbox{ GeV } \,\, .
\end{equation}
This is consistent with our axiflavon constraint in Eq.~\eqref{mesonbound}.

\section{conclusions}  \label{conclusions}
In this paper, we have studied a nonsupersymmetric flavor model based on the double tetrahedral group, $T'$.  Improving on 
earlier work by Carone, Chaurasia and Vasquez~\cite{Carone:2016xsi}, we formulate a simpler model that dispenses with the triplet flavon 
$S$ and eliminates some small numerical coefficients that were assumed in one version of the model to arise from unspecified physics at 
higher energy scales.  Moreover, by replacing one of the Abelian discrete group factors by a continuous U(1) flavor symmetry, we endow the 
theory with a flavorful axion that solves the strong CP problem.  The flavorful axion decay constant $f_a$ is
related to the flavor scale $M_F$ (the cut off of the effective theory) and falls roughly two orders of magnitude beneath it.  We present 
constraints on $f_a$ coming from FCNC processes and find that the strongest lower bound comes from the process 
$K^+ \rightarrow \pi^+a$, yielding $f_a >1.2\times10^{11}$~GeV.   We 
show that the Yukawa matrices predicted by the model provide a good fit to the observed charged fermion masses and mixing angles, taking 
into account the running from the flavor scale down to the weak scale.   We then successfully extend the model to the neutrino sector, by 
introducing three generations of right-handed neutrinos and employing a Type-I see-saw mechanism to explain the smallness of the light 
neutrino masses.  By charging only the first generation 
right-handed neutrino non-trivially under $T'$, we show how the mass matrix for the light neutrino mass eigenstates, which must account for 
two large mixing angles and requires only a modest hierarchy between the neutrino masses, can be predicted by the same theory that yields 
the strong hierarchies of the charged fermion Yukawa matrices.   For the particular extension to the neutrino sector presented here, the flavor 
scale is roughly five orders of magnitude higher than what is required to satisfy the flavorful axion bounds. This suggests that flavor-changing 
signals from the flavorful axion will not be easily observable unless additional symmetries are introduced to lower the scale associated 
with the right-handed neutrinos.

\begin{acknowledgments}  
We thank Tangereen Claringbold for useful discussions and Shikha Chaurasia for cross checking some of our numerical results.
We thank the NSF for support under Grant PHY-1819575.    
\end{acknowledgments}



\begin{thebibliography}{99}

\bibitem{Aranda:1999kc} 
  A.~Aranda, C.~D.~Carone and R.~F.~Lebed,
  ``U(2) flavor physics without U(2) symmetry,''
  Phys.\ Lett.\ B {\bf 474}, 170 (2000)
  [hep-ph/9910392].
\bibitem{Aranda:2000tm} 
  A.~Aranda, C.~D.~Carone and R.~F.~Lebed,
  ``Maximal neutrino mixing from a minimal flavor symmetry,''
  Phys.\ Rev.\ D {\bf 62}, 016009 (2000)
  [hep-ph/0002044].
\bibitem{Barbieri:1995uv} 
  R.~Barbieri, G.~R.~Dvali and L.~J.~Hall,
  ``Predictions from a U(2) flavor symmetry in supersymmetric theories,''
  Phys.\ Lett.\ B {\bf 377}, 76 (1996)
  [hep-ph/9512388].
\bibitem{Barbieri:1996ww} 
  R.~Barbieri, L.~J.~Hall, S.~Raby and A.~Romanino,
  ``Unified theories with U(2) flavor symmetry,''
  Nucl.\ Phys.\ B {\bf 493}, 3 (1997)
  [hep-ph/9610449].
\bibitem{early}
P.~H.~Frampton and T.~W.~Kephart,
  ``Simple nonAbelian finite flavor groups and fermion masses,''
  Int.\ J.\ Mod.\ Phys.\ A {\bf 10}, 4689 (1995)
  [hep-ph/9409330].
 \bibitem{manytp}
  I.~Girardi, A.~Meroni, S.~T.~Petcov and M.~Spinrath,
  ``Generalised geometrical CP violation in a $T'$ lepton flavour model,''
  JHEP {\bf 1402}, 050 (2014)
  [arXiv:1312.1966 [hep-ph]];
  M.~C.~Chen, J.~Huang, K.~T.~Mahanthappa and A.~M.~Wijangco,
  ``Large $\theta_{13}$ in a SUSY SU(5)$\times T'$ Model,''
  JHEP {\bf 1310}, 112 (2013)
  [arXiv:1307.7711 [hep-ph]];
  P.~H.~Frampton, C.~M.~Ho and T.~W.~Kephart,
  ``Heterotic discrete flavor model,''
  Phys.\ Rev.\ D {\bf 89}, no. 2, 027701 (2014)
  [arXiv:1305.4402 [hep-ph]];
   Y.~H.~Ahn,
  ``Leptons and Quarks from a Discrete Flavor Symmetry,''
  Phys.\ Rev.\ D {\bf 87}, no. 11, 113011 (2013)
  [arXiv:1303.4863 [hep-ph]];
  A.~Meroni, E.~Molinaro and S.~T.~Petcov,
  ``Revisiting Leptogenesis in a SUSY SU(5)$\times T'$ Model of Flavour,''
  Phys.\ Lett.\ B {\bf 710}, 435 (2012)
  [arXiv:1203.4435 [hep-ph]];
  D.~A.~Eby and P.~H.~Frampton,
  ``Nonzero $\theta_{13}$ signals nonmaximal atmospheric neutrino mixing,''
  Phys.\ Rev.\ D {\bf 86}, 117304 (2012)
  [arXiv:1112.2675 [hep-ph]];
  D.~A.~Eby and P.~H.~Frampton,
  ``Dark Matter from Binary Tetrahedral Flavor Symmetry,''
  Phys.\ Lett.\ B {\bf 713}, 249 (2012)
  [arXiv:1111.4938 [hep-ph]];
  D.~A.~Eby, P.~H.~Frampton, X.~G.~He and T.~W.~Kephart,
  ``Quartification with $T'$ Flavor,''
  Phys.\ Rev.\ D {\bf 84}, 037302 (2011)
  [arXiv:1103.5737 [hep-ph]];
  M.~C.~Chen, K.~T.~Mahanthappa and F.~Yu,
  ``A Viable Randall-Sundrum Model for Quarks and Leptons with $T'$ Family Symmetry,''
  Phys.\ Rev.\ D {\bf 81}, 036004 (2010)
  [arXiv:0907.3963 [hep-ph]];
  M.~C.~Chen and K.~T.~Mahanthappa,
  ``Group Theoretical Origin of CP Violation,''
  Phys.\ Lett.\ B {\bf 681}, 444 (2009)
  [arXiv:0904.1721 [hep-ph]];
  P.~H.~Frampton, T.~W.~Kephart and S.~Matsuzaki,
  ``Simplified Renormalizable $T'$ Model for Tribimaximal Mixing and Cabibbo Angle,''
  Phys.\ Rev.\ D {\bf 78}, 073004 (2008)
  [arXiv:0807.4713 [hep-ph]];
  C.~Luhn,
  ``Discrete Anomalies of Binary Groups,''
   Phys.\ Lett.\ B {\bf 670}, 390 (2009)
  [arXiv:0807.1749 [hep-ph]];
  G.~J.~Ding,
  ``Fermion Mass Hierarchies and Flavor Mixing from $T'$ Symmetry,''
  Phys.\ Rev.\ D {\bf 78}, 036011 (2008)
  [arXiv:0803.2278 [hep-ph]];
  S.~Sen,
  ``Quark masses in supersymmetric SU(3)$_C \times$ SU(3)$_W \times$ U(1)$_X$ model with discrete $T'$ flavor symmetry,''
  Phys.\ Rev.\ D {\bf 76}, 115020 (2007)
  [arXiv:0710.2734 [hep-ph]];
  A.~Aranda,
  ``Neutrino mixing from the double tetrahedral group $T'$,''
  Phys.\ Rev.\ D {\bf 76}, 111301 (2007)
  [arXiv:0707.3661 [hep-ph]];
  P.~H.~Frampton and T.~W.~Kephart,
  ``Flavor Symmetry for Quarks and Leptons,''
  JHEP {\bf 0709}, 110 (2007)
  [arXiv:0706.1186 [hep-ph]];
  M.~C.~Chen and K.~T.~Mahanthappa,
  ``CKM and Tri-bimaximal MNS Matrices in a $SU(5) \times ^{(d)}\!\!T$ Model,''
  Phys.\ Lett.\ B {\bf 652}, 34 (2007)
  [arXiv:0705.0714 [hep-ph]];
  F.~Feruglio, C.~Hagedorn, Y.~Lin and L.~Merlo,
  ``Tri-bimaximal Neutrino Mixing and Quark Masses from a Discrete Flavour Symmetry,''
  Nucl.\ Phys.\ B {\bf 775}, 120 (2007)
  Erratum: [Nucl.\ Phys.\ B {\bf 836}, 127 (2010)]
  [hep-ph/0702194];
\bibitem{Abel:2015oxa} 
  S.~Abel, K.~R.~Dienes and E.~Mavroudi,
  ``Towards a nonsupersymmetric string phenomenology,''
  Phys.\ Rev.\ D {\bf 91}, no. 12, 126014 (2015)
  [arXiv:1502.03087 [hep-th]].
  \bibitem{Graham:2015cka} 
  P.~W.~Graham, D.~E.~Kaplan and S.~Rajendran,
  ``Cosmological Relaxation of the Electroweak Scale,''
  Phys.\ Rev.\ Lett.\  {\bf 115}, no. 22, 221801 (2015)
  [arXiv:1504.07551 [hep-ph]].
  
  \bibitem{Arkani-Hamed:2016rle} 
  N.~Arkani-Hamed, T.~Cohen, R.~T.~D'Agnolo, A.~Hook, H.~D.~Kim and D.~Pinner,
  ``Solving the Hierarchy Problem at Reheating with a Large Number of Degrees of Freedom,''
  Phys.\ Rev.\ Lett.\  {\bf 117}, no. 25, 251801 (2016)
  [arXiv:1607.06821 [hep-ph]].
  
 \bibitem{Weinberg:1987dv} 
 For example, S.~Weinberg,
  ``Anthropic Bound on the Cosmological Constant,''
  Phys.\ Rev.\ Lett.\  {\bf 59}, 2607 (1987).
  doi:10.1103/PhysRevLett.59.2607
  
  \bibitem{Carone:2016xsi} 
  C.~D.~Carone, S.~Chaurasia and S.~Vasquez,
  ``Flavor from the double tetrahedral group without supersymmetry,''
  Phys.\ Rev.\ D {\bf 95}, no. 1, 015025 (2017)
  [arXiv:1611.00784 [hep-ph]].
   
  \bibitem{Bjorkeroth:2018dzu} 
  F.~Björkeroth, E.~J.~Chun and S.~F.~King,
  ``Flavourful Axion Phenomenology,''
  JHEP {\bf 1808}, 117 (2018)
  [arXiv:1806.00660 [hep-ph]].
  
  \bibitem{Ema:2016ops} 
  Y.~Ema, K.~Hamaguchi, T.~Moroi and K.~Nakayama,
  ``Flaxion: a minimal extension to solve puzzles in the standard model,''
  JHEP {\bf 1701}, 096 (2017)
  [arXiv:1612.05492 [hep-ph]].
 
  \bibitem{Calibbi:2016hwq} 
  L.~Calibbi, F.~Goertz, D.~Redigolo, R.~Ziegler and J.~Zupan,
  ``Minimal axion model from flavor,''
  Phys.\ Rev.\ D {\bf 95}, no. 9, 095009 (2017)
  [arXiv:1612.08040 [hep-ph]].
 
  \bibitem{Linster:2018avp} 
  M.~Linster and R.~Ziegler,
  ``A Realistic $U(2)$ Model of Flavor,''
  JHEP {\bf 1808}, 058 (2018)
  [arXiv:1805.07341 [hep-ph]].
  
 \bibitem{Arias-Aragon:2017eww} 
  F.~Arias-Aragon and L.~Merlo,
  ``The Minimal Flavour Violating Axion,''
  JHEP {\bf 1710}, 168 (2017)
  [arXiv:1709.07039 [hep-ph]].
  
  \bibitem{Ahn:2018cau} 
  Y.~H.~Ahn,
  ``Compact model for Quarks and Leptons via flavored-Axions,''
  Phys.\ Rev.\ D {\bf 98}, no. 3, 035047 (2018)
  [arXiv:1804.06988 [hep-ph]].
 
  \bibitem{Falkowski:2015zwa} 
  A.~Falkowski, M.~Nardecchia and R.~Ziegler,
  ``Lepton Flavor Non-Universality in B-meson Decays from a U(2) Flavor Model,''
  JHEP {\bf 1511}, 173 (2015)
  [arXiv:1509.01249 [hep-ph]].

\bibitem{Krauss:1988zc} 
  L.~M.~Krauss and F.~Wilczek,
  ``Discrete Gauge Symmetry in Continuum Theories,''
  Phys.\ Rev.\ Lett.\  {\bf 62}, 1221 (1989).

 \bibitem{Tanabashi:2018oca} 
  M.~Tanabashi {\it et al.} [Particle Data Group],
  ``Review of Particle Physics,''
  Phys.\ Rev.\ D {\bf 98}, no. 3, 030001 (2018).

\bibitem{Kim:1986ax} 
  J.~E.~Kim,
  ``Light Pseudoscalars, Particle Physics and Cosmology,''
  Phys.\ Rept.\  {\bf 150}, 1 (1987).
   
\bibitem{Marsh:2015xka} 
  D.~J.~E.~Marsh,
  ``Axion Cosmology,''
  Phys.\ Rept.\  {\bf 643}, 1 (2016)
  [arXiv:1510.07633 [astro-ph.CO]].

\bibitem{Feng:1997tn} 
  J.~L.~Feng, T.~Moroi, H.~Murayama and E.~Schnapka,
  ``Third generation familons, b factories, and neutrino cosmology,''
  Phys.\ Rev.\ D {\bf 57}, 5875 (1998)
  [hep-ph/9709411].
   
\bibitem{Bauer:2016rxs} 
  M.~Bauer, T.~Schell and T.~Plehn,
  ``Hunting the Flavon,''
  Phys.\ Rev.\ D {\bf 94}, no. 5, 056003 (2016)
  [arXiv:1603.06950 [hep-ph]].
  
\bibitem{Celis:2014iua} 
  A.~Celis, J.~Fuentes-Martin and H.~Serodio,
  ``An invisible axion model with controlled FCNCs at tree level,''
  Phys.\ Lett.\ B {\bf 741}, 117 (2015)
  [arXiv:1410.6217 [hep-ph]].
 
\bibitem{Albrecht:2010xh} 
  M.~E.~Albrecht, T.~Feldmann and T.~Mannel,
  ``Goldstone Bosons in Effective Theories with Spontaneously Broken Flavour Symmetry,''
  JHEP {\bf 1010}, 089 (2010)
  [arXiv:1002.4798 [hep-ph]].
  
\bibitem{Davidson:1984ik} 
  A.~Davidson and M.~A.~H.~Vozmediano,
  ``The Horizontal Axion Alternative: The Interplay of Vacuum Structure and Flavor Interactions,''
  Nucl.\ Phys.\ B {\bf 248}, 647 (1984).
  
  \bibitem{Adler:2008zza} 
  S.~Adler {\it et al.} [E949 and E787 Collaborations],
  ``Measurement of the $K^+ \rightarrow \pi^+ \nu \nu$ branching ratio,''
  Phys.\ Rev.\ D {\bf 77}, 052003 (2008)
  [arXiv:0709.1000 [hep-ex]].
  
\bibitem{Ahn:2016kja} 
  J.~K.~Ahn {\it et al.} [KOTO Collaboration],
  ``A new search for the $K_{L} \to \pi^0 \nu \overline{\nu}$ and $K_{L} \to \pi^{0} X^{0}$ decays,''
  PTEP {\bf 2017}, no. 2, 021C01 (2017)
  [arXiv:1609.03637 [hep-ex]].
  
\bibitem{Ammar:2001gi} 
  R.~Ammar {\it et al.} [CLEO Collaboration],
  ``Search for the familon via $B^{+-} \rightarrow \pi^{\pm} X^0$, $B^{\pm} \rightarrow K^{\pm} X^0$, and $
  B^0 \rightarrow K^0_S X^0$ decays,''
  Phys.\ Rev.\ Lett.\  {\bf 87}, 271801 (2001)
  [hep-ex/0106038].

\bibitem{Bayes:2014lxz} 
  R.~Bayes {\it et al.} [TWIST Collaboration],
  ``Search for two body muon decay signals,''
  Phys.\ Rev.\ D {\bf 91}, no. 5, 052020 (2015)
  [arXiv:1409.0638 [hep-ex]].
 
\bibitem{Albrecht:1995ht} 
  H.~Albrecht {\it et al.} [ARGUS Collaboration],
  ``A Search for lepton flavor violating decays $\tau \rightarrow e \,\, \alpha, \tau \rightarrow \mu \,\,  \alpha$,''
  Z.\ Phys.\ C {\bf 68}, 25 (1995).


  
  \end{thebibliography}
\end{document}